\documentclass{article}
\usepackage{waspaa23,amsmath,graphicx,url,times}
\usepackage{color}
\usepackage{times}  
\usepackage{helvet} 
\usepackage{courier} 
\usepackage{siunitx}
\usepackage{amsfonts}
\usepackage{algorithm}
\usepackage{algorithmic}
\usepackage{amssymb}
\usepackage{booktabs, tabularx, colortbl}
\usepackage{multirow}
\usepackage{makecell}
\usepackage{amsmath}
\usepackage{subfig}
\usepackage{xcolor}

\usepackage{pifont}
\usepackage{hyperref}
\usepackage{enumitem}
\usepackage{setspace}
\usepackage{lineno}
\usepackage{ifthen}

\title{
Yet Another Generative Model For Room Impulse Response Estimation
}

\name{Sungho Lee$^{1*}$, 
      Hyeong-Seok Choi$^{4\dagger }$,
      and Kyogu Lee$^{1, 2, 3, 4\dagger }$}
\address{$^1$Department of Intelligence and Information, 
        $^2$IPAI,  
        $^3$AI Institute, Seoul National University\\
        $^4$Supertone, Inc.\\
        \texttt{sh-lee@snu.ac.kr, kekepa15@supertone.ai, kglee@snu.ac.kr}
         \thanks{
         \hspace{-5.9mm}
         $^*$Work done during an internship at Supertone, Inc. $^\dagger$Corresponding authors. \\ Project page: \href{https://sh-lee97.github.io/neural-ir-est}{\texttt{https://sh-lee97.github.io/neural-ir-est}}. \\ This work was supported by the Institute of Information \& Communications Technology Planning \& Evaluation (IITP) grant funded by the Korean government (MSIT) (No.2022-0-00641). 
         }
}

\begin{document}

\ninept
\maketitle

\begin{sloppy}

\begin{abstract}
Recent neural room impulse response (RIR) estimators typically comprise an encoder for reference audio analysis and a generator for RIR synthesis. Especially, it is the performance of the generator that directly influences the overall estimation quality. In this context, we explore an alternate generator architecture for improved performance. We first train an autoencoder with residual quantization to learn a discrete latent token space, where each token represents a small time-frequency patch of the RIR. Then, we cast the RIR estimation problem as a reference-conditioned autoregressive token generation task, employing transformer variants that operate across frequency, time, and quantization depth axes. This way, we address the standard blind estimation task and additional acoustic matching problem, which aims to find an RIR that matches the source signal to the target signal's reverberation characteristics. Experimental results show that our system is preferable to other baselines across various evaluation metrics.
\end{abstract}

\vspace{-2mm}
\section{Introduction}
\vspace{-2.5mm}

\noindent \textbf{Motivation.}
If a signal processing system is linear time-invariant (LTI), we can fully describe its input-output relationship with a signal called impulse response (IR). In particular, room impulse responses (RIRs), which represent sound propagation in real-world environments, have been used for the analysis of acoustic properties \cite{2008ISO3382_2}, auralization of dry signals \cite{steinmetz21fins, lee2022differentiable}, improving speech recognition models \cite{ratnarajah2023towards}, and much more. While one can directly measure RIRs with dedicated equipment, such a process is often time-consuming and costly. 
Therefore, there is a compelling need for models that can estimate RIRs from \emph{wet} signals such as reverberant speech.

\noindent \textbf{Related Works.}
Several neural networks for RIR estimation have been recently proposed \cite{steinmetz21fins, lee2022differentiable, ratnarajah2023towards}. 
These models typically consist of two components: (i) an encoder that extracts RIR-relevant features from reference and (ii) a generator that synthesizes the RIR from the extracted features. 
Here, the generator's architecture is where these methods differ the most. 
For example, Filtered Noise Shaping (FiNS) \cite{steinmetz21fins} outputs attenuation gain envelope of trained filterbank noise with additional early reflection signal. 
This design choice stems from the observation that late reverberation exhibits perceptual similarities to time-varying filtered noise. 
While FiNS offers a balanced approach, further inductive bias can be introduced to the generator; we can use an artificial reverberation algorithm as a (trainable parameter-less) RIR generator \cite{lee2022differentiable}. 
This approach offers benefits like training data efficiency and computation efficiency during auralization. 
However, as more RIR data \cite{simon2010openair, traer2016reverbperception, eaton2016ace, Thomas2021coupledroom} and computation budget become available, compute/data-demanding but more expressive neural architecture might be desirable. 
In this spirit, we could remove the filtered noise component from the FiNS and develop a more general and flexible generator, like the speech-to-IR generative adversarial network (S2IR-GAN) \cite{ratnarajah2023towards}, which uses stacked convolutional layers for the RIR synthesis. 
In fact, we could borrow any neural network developed for audio synthesis. 
However, RIRs (and IRs in general) exhibit unique signal characteristics, such as sparsity and sharp transients; hence, neural networks suitable for RIR generation might be different and remain to be explored.

\noindent \textbf{Contributions.}
In this paper, we explore an alternative generator architecture and RIR estimation framework (See Figure \ref{fig:framework}). 
First, a residual-quantized variational autoencoder (RQ-VAE) \cite{zeghidour2021soundstream} is trained to reconstruct the time-frequency features of RIRs, learning the discrete intermediate representation space. With the trained RQ-VAE, we convert each RIR into discrete token indices (or an index tensor). 
Then, we introduce transformer-based models \cite{vaswani2017attention} that decode the token indices autoregressively. 
Each RIR index tensor is three-dimensional with time, frequency, and quantization depth axis; inspired by recent works \cite{borsos2022audiolm, lee2022autoregressive, wang2023neural}, we explore various strategies to decode this tensor in a hierarchical and factorized manner.
Finally, by encoding the references and providing the features as \emph{prompt} tokens to the transformers, we cast the RIR estimation problem to a conditional generative modeling task.
Along with the standard blind estimation (reverberant-speech-to-RIR), we also address the acoustic matching scenario, whose goal is to find an RIR that matches the source signal to the target signal's reverberation characteristics.
The evaluation results show that our system can estimate perceptually close RIRs.
Also, the ablation study indicates that both discretization and autoregression are essential for plausible performance. 

\vspace{-2.5mm}
\section{Proposed Framework}
\vspace{-2mm}
\subsection{Discrete Representation Learning with RQ-VAE}
\vspace{-1.5mm}
\noindent \textbf{Input Representation.} 
We first convert a single-channel RIR $h\in \mathbb{R}^{49152}$ ($1.1$ second, $44.1\si{kHz}$ sampling rate) into a time-frequency representation with a short-time Fourier transform (STFT). We use relatively short FFT and hop size, $512$ and $64$, to increase the time-axis resolution and capture the transient/sparse region. We also apply power compression \cite{braun2021consolidated} to the STFT. 
We stack the magnitude, real, and imaginary spectrograms and obtain a three-channel feature $\mathbf{H}\in\mathbb{R}^{3\times 256\times 768}$ ($256$ frequency bins and $768$ time frames). 

\noindent \textbf{Autoencoding.} 
We compress and expand the feature $\mathbf{H}$ using an autoencoder. Our encoder follows conventional architectures \cite{van2017neural, esser2021taming}, stacking the residual blocks and downsampling layers. We observed that, however, downsampling the frequency axis with strided convolutions compresses the low-frequency region too much. To mitigate this, we replace them with nonlinear filterbank matrices; each matrix halves the frequency dimension, and the cascade of the $3$ filterbanks gives a frequency mapping that resembles the Mel scale. We downsample the time axis with strided convolutions. We reduce the dimension of frequency and time axis by $8$ and $4$ times and increase the channels to $192$, resulting in a latent $\mathbf{S} \in \mathbb{R}^{192\times 32\times 192}$. The decoder mirrors the encoder with transposed filterbanks/convolutions.

\noindent \textbf{Residual Quantization.} 
Then, we discretize the latent $\mathbf{S}$ with residual quantization (RQ), i.e., we iteratively quantize the residual error of vector quantization \cite{van2017neural, juang1982multiple}. We use $3$ codebooks, each with $512$ codes; we found that deeper codebook depth than this is redundant. We update each codebook with exponential moving average with a decay of $0.98$. 
We initialize the codes with k-means clustering, re-initialize the stale codes, and use quantization dropout during training \cite{zeghidour2021soundstream}. 
This way, we convert each RIR $h$ into a quantized feature $\mathbf{S}^\mathrm{q} \in \mathbb{R}^{192\times 32\times 192}$ or its corresponding indices $\mathbf{Q} \in \mathbb{N}^{32\times 192\times 3}$. 

\noindent \textbf{Differences to Existing Neural Audio Codecs.} Our RQ-VAE is similar to existing neural audio codecs, e.g., SoundStream \cite{zeghidour2021soundstream} and Encodec \cite{defossez2022high}, as it employs residual quantization, but has two major differences. 
First, each frequency band is encoded into separate tokens $\mathbf{Q}_f$\footnote{\textbf{Notation.} We denote a single index with frequency $f$, time $t$, and quantizer depth $d$ as $\mathbf{Q}_{f, t, d}$. We omit the subscripts if we consider an index vector or tensor, e.g., $\mathbf{Q}_{f}\in \mathbb{R}^{192\times 4}$ denotes the codes with $f^\text{th}$ frequency bin. $\mathbf{Q}_{<t}$ denotes codes with a time index less than $t$.}, as an LTI system can have arbitrary responses in one band regardless of the others.
Another difference is that our objective is not to compress the input data in terms of bitrate. Instead, our RQ-VAE's encoder layer is closer to the \emph{patchify} layer for the vision transformer \cite{dosovitskiyimage}, whose role is to embed the data into input tokens; each latent vector $\mathbf{S}_{f, t} \in \mathbb{R}^{192}$ (or its discrete tokens $\mathbf{Q}_{f, t}\in \mathbb{N}^{3}$) represents an overlapping time-frequency patch.

\vspace{-3mm}
\subsection{Autoregressive Token Generation with Axial Transformers}
\vspace{-1.5mm}

\noindent \textbf{Factorization Strategies.} 
Since we discretized each RIR $h$, our objective is to generate the index tensor $\mathbf{Q}$, which is three-dimensional with frequency, time, and depth axis. We want to apply autoregressive generation, but flattening all three axes results in a prohibitively long sequence (length $18432$). Instead, we decode it in a more factorized way. First, we treat each frequency column $\mathbf{Q}_{t, d}\in\mathbb{N}^{32}$ as a single decoding unit. Then, as the time and depth axis still remain, we decode them with the following three different strategies. 
\setlength{\belowdisplayskip}{4.4pt} \setlength{\belowdisplayshortskip}{4.4pt}
\setlength{\abovedisplayskip}{4.4pt} \setlength{\abovedisplayshortskip}{4.4pt}

\vspace{-1.5mm}
\begin{enumerate}[leftmargin=1\parindent]
    \item We flatten the remaining axes, then decode each column autoregressively across the flattened axis (see Figure \ref{fig:audiolm}). In this way, we model the probability distribution of the indices as follows,
    \begin{equation}
    p(\mathbf{Q}) = \prod_{t, d, f}  p(\mathbf{Q}_{f, t, d}| \mathbf{Q}_{<t}, \mathbf{Q}_{t, <d}).
    \end{equation}
    Implementation-wise, we start with a start-of-sequence column $\mathbf{x}_\mathrm{S} \in \mathbb{R}^{384\times 32\times 1}$. We pass it to a Time-Frequency Transformer (TF-Transformer), a transformer variant model, to obtain a feature with a shape the same as the input. With a linear projection, we predict the first column $\mathbf{Q}_{0,0} \in \mathbb{N}^{32}$ of the index tensor. The predicted indices are converted back to the latent with the codebook lookup and preprocessed with a linear layer, resulting in a transformer input $\mathbf{x}_{0, 0} \in \mathbb{R}^{384\times 32\times 1}$. We concatenate it with the start-of-sequence $\mathbf{x}_\mathrm{S} | \mathbf{x}_{0, 0}$, pass to the TF-Transformer, then predict the next column $\mathbf{Q}_{0, 1}$. We repeat this autoregressively, decoding $\mathbf{Q}_{0, 2}$, $\mathbf{Q}_{1, 0}$, $\mathbf{Q}_{1, 1}$, and so on, until the model predicts an end-of-sequence. 
    This approach resembles AudioLM \cite{borsos2022audiolm} if we ignore that we generate a frequency column every time.
    
    \item While keeping the previous decoding order, we can use separate models to decode across time and depth (see Figure \ref{fig:rqt}). 
    \begin{equation}
    p(\mathbf{Q}) =  \prod_{t, f}  p(\mathbf{Q}_{f, t, 0}| \mathbf{Q}_{<t}) \prod_d  p(\mathbf{Q}_{f, t, d}| \mathbf{Q}_{f,t,<d}, \mathbf{Q}_{<t}).
    \end{equation}
    Specifically, a TF-Transformer processes a partially decoded latent $\mathbf{S}^\mathrm{q}_{<t}\in \mathbb{R}^{384\times 32\times (t-1)}$. Each output feature $\mathbf{o}_{f, t} \in \mathbb{R}^{384}$ is then passed to another transformer called Depth Transformer to independently and autoregressively estimate the indices $\mathbf{Q}_{f, t, 0}$, $\mathbf{Q}_{f,t, 1}$, and $\mathbf{Q}_{f, t, 2}$. This setup reduces the maximum input sequence length of the TF-Transformer from $576$ to $192$. Instead, the more lightweight Depth Transformer is evaluated $576$ times, reducing the total computation/memory usage.
    Except for the frequency axis, this approach resembles RQ-Transformer \cite{lee2022autoregressive}.
    
    \item We can further reduce the number of forward passes by borrowing Vall-E's approach \cite{wang2023neural}. Instead of interleaving the time and depth axis, we first predict the depth-$0$ indices $\mathbf{Q}_0$, then predict the following depths, $\mathbf{Q}_1$ and $\mathbf{Q}_2$, with $2$ passes (see Figure \ref{fig:valle}).
    \begin{equation}
        p(\mathbf{Q}) =  \prod_{t, f}  p(\mathbf{Q}_{f, t, 0}| \mathbf{Q}_{<t, 0}) \prod_d p(\mathbf{Q}_{d}|\mathbf{Q}_{<d}) .
    \end{equation}
    The depth-$0$ decoding is identical to the first AudioLM-like approach. The decoded tokens $\mathbf{S}_0\in\mathbb{R}^{384\times 32\times 192}$ are then passed to another TF-Transformer without the causal mask to estimate the following $\mathbf{Q}_1$ and $\mathbf{Q}_2$. We additionally append a depth embedding to the input of the second transformer.
\end{enumerate}

\begin{figure}
    \begin{center}
    \includegraphics[width=.98\columnwidth]{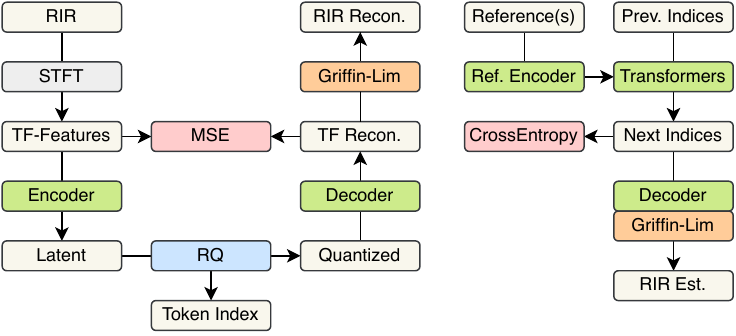} \vspace{-2mm}
        \caption{The proposed method (left: codebook learning with RQ-VAE, right: RIR estimation via conditional token generation). } 
        \label{fig:framework}
    \end{center}
    \vspace{-3mm}
\end{figure}
\begin{figure}
    \centering
    \subfloat[Autoregression across flattened time/depth ($\approx$ AudioLM). \label{fig:audiolm}]{
        \includegraphics[width=.98\columnwidth]{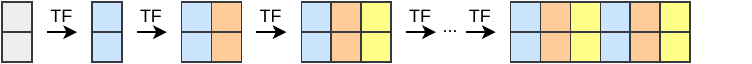}}\vspace{-2mm} \\
    \subfloat[Factorized autoregression across time and depth ($\approx$ RQ-Transformer). \label{fig:rqt}]{
        \includegraphics[width=.98\columnwidth]{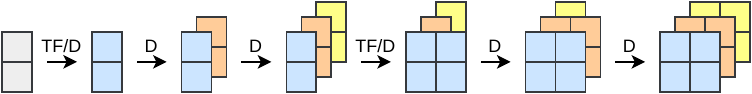}}\vspace{-2mm} \\
    \subfloat[Factorized autoregression of depth $0$ across time, followed by full time-frequency token generation across depth ($\approx$ VALL-E). \label{fig:valle}]{\vspace{0.3mm}
        \includegraphics[width=.98\columnwidth]{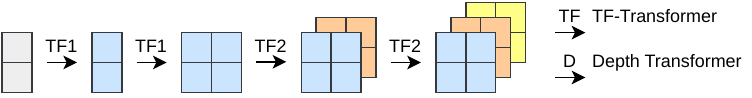}}\vspace{-2mm} \\
\caption{Token generation strategies with axial transformers.}
\vspace{-4mm}
  \label{fig:properties} 
\end{figure}

\noindent \textbf{Transformer Architecture.} 
    The TF-Transformer is a stack of four blocks composed of a frequency-axis encoder layer and a time-axis decoder-only layer, which treat the time and frequency axis as batch axis, respectively. In this setup, the former exchanges information between different frequency bands, while the latter autoregressively processes the feature across the time (and optional flattened depth).
    The Depth Transformer is a two-layer decoder-only model. All layers have pre-layer normalization \cite{xiong2020layer}, $384$/$1536$ input/feed-forward dimension, $4$ heads, and dropout of $0.1$.
    
\noindent \textbf{Inference.} 
For the inference-time generation, we employ top-$k$ and top-$p$ sampling, where the hyperparameters are set to $k=10$ and $p=0.995$.
The decoded indices are converted back to the features, then fed into the frozen RQ-VAE decoder to obtain the spectrogram $\mathbf{H}$.
We convert the spectrogram back to the time-domain RIR $h$ with the Griffin-Lim algorithm \cite{griffin1984signal}; we provide the magnitude with the initial phase obtained from the real/imaginary part.

\vspace{-3mm}
\subsection{Conditional Generation with Reference Encoder}
\vspace{-1mm}
\noindent \textbf{Reference Encoder.} 
By providing conditioning information to the transformers, we can cast various RIR estimation problems as conditional generative modeling. For example, if we provide an RIR $h$ as a reference, the task becomes \emph{analysis-synthesis}. If reverberant speech $y=h*x$ is given, we tackle \emph{blind estimation}. 
To condition the reference, we first encode it 
with a reference encoder. The blind estimation (and the following acoustic matching) task requires us to extract reverberation from the speech signal; hence, we borrow the network architectures from speech enhancement literature. Specifically, we use the encoder (downsampling) part of MFTAA-Net \cite{zhang2022multi}, which stacks frequency-axis downsampling layers, time-frequency dilated convolutional layers and time-frequency axial self-attention layers. We modify this model by replacing the downsampling layers with the nonlinear filter bank matrices similar to the RQ-VAE and doubling the number of channels. To the output of the modified MFTAA encoder, we apply attentive pooling across the time axis with $4$ learnable queries, resulting in a latent tensor $\mathbf{z} \in \mathbb{R}^{384\times 32\times 4}$, i.e., $4$ latent vectors $\mathbf{z}_f \in \mathbb{R}^{384\times 4}$ for each frequency bin.
To condition the latent, we simply concatenate it to the TF-Transformer input. 
As the reference encoder's frequency mapping matches the RQ-VAE's, the $f^\text{th}$ band latent $\mathbf{z}_f$ is used to estimate the corresponding band's indices $\mathbf{Q}_f \in \mathbb{N}^{32\times 192}$. 
Such a design choice aligns with the property of LTI systems; 
they process each frequency band of input with its corresponding band's frequency response. 

\noindent \textbf{Acoustic Matching.} 
Here, we further tackle an \emph{acoustic matching} problem; we predict the RIR $h$ from a dry source $x^\mathrm{s}$ and wet target $y^\mathrm{t}=h*x^\mathrm{t}$, where the source and target's dry signals are not necessarily the same. We encode both signal with the reference encoder and subtract the two to obtain the latent $\mathbf{z} = \mathbf{z}^\mathrm{t}-\mathbf{z}^\mathrm{s}$, which conditions the transformers. This way, we train our models to relatively match the source to the target. We find that the acoustic matching setup is more practical than blind estimation, as a typical application of RIR estimation is to apply the estimation to another dry signal.

\vspace{-2.5mm}
\section{Experiments}
\vspace{-3mm}
\subsection{Data}
\vspace{-1.5mm}
\noindent \textbf{Room Impulse Responses.} 
We combined multiple public datasets \cite{simon2010openair, traer2016reverbperception, eaton2016ace, Thomas2021coupledroom, szoke2019building, yasuda2022echo, stewart2010database, amengual2020open, kearney2022measuring, dietzen2023myriad, nakamura1999sound, jeub2009binaural} 
\footnote{We additionally used \href{http://www.echothief.com}{Echotheif}, \href{https://fokkie.home.xs4all.nl/IR.htm}{Fokke}, and \href{https://research.kent.ac.uk/sonic-palimpsest/impulse-responses}{Palimpsest} RIR database.}. All of the RIRs are from real spaces, ranging from daily environments, e.g., living rooms, to musical ones, e.g., concert halls.
We collected $15\si{k}$ RIRs from $559$ rooms 
and split them into $496$/$16$/$47$ for the training/validation/test. 

\noindent \textbf{Microphone Impulse Responses.} 
We used additional microphone IRs (MicIRs) \cite{franco2022multi} \footnote{We also included  \href{https://micirp.blogspot.com}{Vintage MicIRs.}} to mimic the colored magnitude responses of real-world recordings. 
We collected a total of $8.4\si{k}$ MicIRs from $82$ microphones and split them into $72$/$3$/$7$. Figure \ref{fig:stats} reports reverberation parameter statistics of the collected RIR/MicIR data.

\begin{figure}
    \begin{center}
    \includegraphics[width=.96\columnwidth]{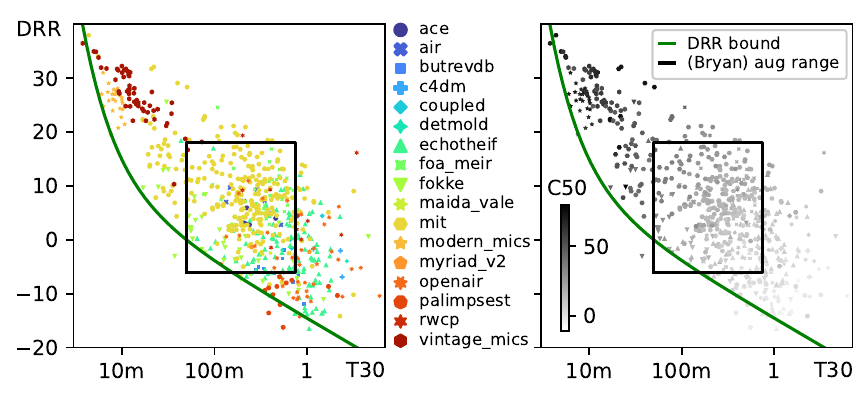} \vspace{-2mm}
        \caption{Reverberation parameter statistics of the IR datasets.} 
        \label{fig:stats}
    \end{center}
    \vspace{-8.5mm}
\end{figure}

\noindent \textbf{Data Augmentation.}
Due to the small amount of RIR/MicIR data, we used various augmentations to mitigate overfitting. We (i) modified the $T_{30}$ and DRR by windowing/scaling \cite{bryan20riraug}.
Then, we also (ii) shifted the pitch with resampling, (iii) applied a parametric equalizer to transform the overall magnitude, and (iv) mixed the channels to mono for ambisonic IRs. We applied the same augmentation procedure to both RIRs and MicIRs, but the hyperparameters were different. Furthermore, we (v) convolved an (augmented) RIR $h^\mathrm{r}$ with a MicIR $h^\mathrm{m}$ to obtain another IR $h^\mathrm{r}*h^\mathrm{m}$ and used it for (target) reverberant speech $y=h^\mathrm{r}*h^\mathrm{m}*x$. For the acoustic matching task, we optionally (vi) applied the MicIR $h^\mathrm{m}$ to source speech to make the distribution of dry signal diverse. In this case, we trained the models to estimate only the RIR $h^\mathrm{r}$. All augmentations are done on the fly, and we randomly choose whether to apply each of them. 

\noindent \textbf{Dry Speech.} We used DAPS \cite{mysore15daps} for the blind estimation models; we split the speakers into $16$/$2$/$2$ for the training/evaluation/test. For acoustic matching, we ensembled more datasets, including NUS \cite{duan2013nus} and VCTK \cite{veaux2017cstr}. 
Some datasets use a single fixed recording env and have consistent acoustic characteristics across the speakers.
For them, we allowed the source and target to be sampled from different speakers. In other cases, we sampled a source/target pair from the same speaker. We split the speakers into $3972$/$51$ for the training/validation and used the NUS solely for the test. We used $3$ and $12$ seconds of dry signals for training and evaluations, respectively.

\vspace{-3mm}
\subsection{Metric}
\vspace{-2mm}
\noindent \textbf{Objective Metrics.} 
We calculate the log spectral distance (LSD), a mean absolute error (MAE) of log-magnitude response. We also report MAE of energy decay relief (EDR) \cite{jot92fdnanalysissynthesis} and multi-scale spectral (MSS) loss \cite{lee2022differentiable, steinmetz21fins}, which is MAE of STFT magnitudes with multiple FFT sizes. Furthermore, we evaluate differences in reverberation time $T_{30}$, direct-to-reverberant ratio (DRR), and clarity $C_{50}$. We averaged differences at octave bands (center frequencies from $125\si{Hz}$ to $8\si{kHz}$), then reported the median of the total results.

\noindent \textbf{Subjective Listening Test.}
We measured subjective reverberation similarity scores (SUBJ) with MUltiple Stimuli with Hidden Reference and Anchor (MUSHRA)-like test \cite{mushra}. 
We asked $8$ subjects with critical listening experiences to rate the similarity between the reference reverberant speech and the dry speech convolved with the estimated IRs. 
We included an average of test RIRs as a low anchor.  After $2$ training sessions, a total of $10$ sets were scored.

\vspace{-3mm}
\subsection{Training}
\vspace{-2mm}
We trained the RQ-VAE with mean square error, where we weighed the magnitude and the complex STFT to $3\!:\!1$. 
We trained the token estimation stage with the cross-entropy loss with label smoothing of $0.001$. We used AdamW \cite{loshchilov2017decoupled} optimizer, $5\si{k}$ linear warm-up steps with \texttt{1e-4} peak learning rate, and halved the learning rate for every $50\si{k}$ steps. We set the batch size to $16$ and $6$ for the first and the second stage. We finished the training when the validation metric did not improve over several epochs. 

\vspace{-3.5mm}
\subsection{Evaluation Results}
\vspace{-1.5mm}
\noindent \textbf{Autoencoding.} 
Table \ref{table:results} reports the evaluation results. 
We first assessed the autoencoder model's performance with and without the quantization layer to understand its impact. Without quantization, the reconstruction is nearly perfect, showing $0.54\si{dB}$ LSD and $2.2\%$ $T_{30}$ error. Adding the vector quantization layer increases these errors to $1.85\si{dB}$ and $9.6\%$, which may be noticeable. However, the residual quantization consistently improves all metrics and achieves a SUBJ score of $87.6_{\pm 4.3}$, close to the hidden reference's $90.5_{\pm 3.1}$. These results are the performance upper bounds of the transformer models as they rely on the RQ-VAE indices and decoder.

\noindent \textbf{Analysis-Synthesis.} 
For sanity-check purposes, we trained the proposed transformer models to perform analysis-synthesis; (i) the reference encoder should provide the conditioning latent sufficient for the reconstruction, and (ii) the transformers should decode the RIR tokens from the latent accurately. Without surprise, these models report slightly worse results than the RQ-VAE, which has access to the ground-truth tokens. Nevertheless, the reconstruction is perceptually similar; the AudioLM-like showed a SUBJ score of $84.0_{\pm 4.0}$. 

\noindent \textbf{Blind Estimation.} 
Estimating RIRs from the reverberant speech is more challenging than analysis-synthesis, and the evaluation results indicate that the estimation error could be perceptually noticeable in terms of reverberation parameters. For example, all three transformer models report $T_{30}$ error larger than $12\%$. The DRR errors also increase significantly from the previous analysis-synthesis results (maximum $2.27\si{dB}$ increase). 
However, the LSD increases by only about $0.5\si{dB}$, indicating that our models can still accurately capture the overall time-averaged characteristics.

\noindent \textbf{Acoustic Matching.} Acoustic matching is a relaxation of blind estimation as it provides the dry source; 
at the same time, it is an extension of the latter because it should accept the sources with arbitrary acoustic conditions, e.g., microphones, which makes the task challenging. Indeed, the acoustic matching models report slightly worse results than the blind estimation counterparts, although the differences are not large, e.g., at most $0.5\si{dB}$ increase in EDR error. 
Furthermore, we analyzed whether providing different speakers for the source/target affects the performance. As shown in the AudioLM-like model's results, the different-speaker scenario reports slightly higher errors. For the other models, we report the aggregated objective metrics of the same-speaker and different-speaker results, while the SUBJ scores are different-speaker results.

\noindent \textbf{Decoding Strategies.}
In general, the heavier methods performed better, e.g., the AudioLM-like showed better results than the VALL-E-like in most objective metrics, but the differences were relatively small, especially in the acoustic matching scenario. Also, the SUBJ scores indicate that such differences might not be perceptually significant ($66.2_{\pm 6.3}$, $70.4_{\pm 5.1}$, and $66.1_{\pm 6.0}$). This might be because the reference encoder provided similar noisy estimates to the transformer models. Moreover, the AudioLM-like took inference time approximately $3.5$ and $6.2$ times more than the RQ-Transformer-like and VALL-E-like, respectively. Therefore, the optimal model choice could differ according to the available computation budget. 
\begin{table}[t]
\setlength\tabcolsep{1.96pt}
\renewcommand{\arraystretch}{.78}
\begin{center}
\fontsize{8.1}{8.1}\selectfont
\begin{tabular}{l|rrrrrr|c}
\toprule
%& & \multicolumn{7}{c}{Singing voice effect estimation} \\
& \multicolumn{6}{c|}{Objective metrics ($\downarrow$)} & \multirow{2}[2]{*}{SUBJ ($\uparrow$)}
\\
\cmidrule{2-7}

Methods & 
\iffalse
\mcrot{1}{l}{27}{Magnitude response \!\!\!} &
\mcrot{1}{l}{27}{Multi-scale STFT \!} &
\mcrot{1}{l}{27}{Energy decay relief \!\!\!} &
\mcrot{1}{l}{27}{Reverberation time \!} &
\mcrot{1}{l}{27}{Direct-to-reverb. ratio \!\!\!} &
\mcrot{1}{l}{27}{Clarity \!} &
MUSHRA
\fi
LSD & MSS & EDR & $T_{30}$ & DRR & $C_{50}$ &  
\\
%& MUSHRA  \\
\midrule

Hidden reference
& $-$    & $-$    & $-$    & $-$    & $-$    & $-$  & $90.5_{\pm 3.1}$\\
Average test RIR
& $11.3$ & $.206$ & $19.4$ & $4.94$  & $9.21$    & $13.4$  & $14.9_{\pm 4.0}$\\

\midrule
\multicolumn{8}{r}{\emph{Autoencoding results}} \\
\midrule

No quantization 
& $0.54$ & $.010$ & $0.61$ & $.022$ & $0.39$ & $0.35$ & $-$ \\
Vector quantization 
& $1.85$ & $.050$ & $1.48$ & $.096$ & $1.63$ & $1.37$ & $-$ \\

Residual quantization
& $1.13$ & $.025$ & $0.76$ & $.034$ & $0.66$ & $0.59$ & $87.6_{\pm 4.3}$\\
\midrule
\multicolumn{8}{r}{\emph{Analysis-synthesis results}} \\
\midrule

AudioLM-like
& ${2.15}$ & $.080$ & ${1.75}$ & $.063$ & ${1.45}$ & $0.97$ & $84.0_{\pm 4.0}$\\
RQ-Transformer-like
& $2.31$ & ${.078}$ & $1.82$ & ${.061}$ & $1.51$ & $1.05$ & $-$ \\
VALL-E-like
& $2.28$ & $.083$ & $1.84$ & $.074$ & $1.58$ & $1.23$ & $-$ \\
Differentiable FVN
& $5.95$ & $.098$ & $1.88$ & $.124$ & $6.53$ & $1.27$ & $-$ \\

\midrule
\multicolumn{8}{r}{\emph{Blind estimation results}} \\
\midrule
AudioLM-like
& $2.67$ & $.122$ & $3.48$ & $.121$ & $3.25$ & $1.69$ & $74.9_{\pm 5.1}$\\
RQ-Transformer-like 
& $2.84$ & $.120$ & $3.55$ & $.123$ & $3.78$ & $1.63$ & $-$ \\
VALL-E-like 
& $2.83$ & $.117$ & $3.77$ & $.149$ & $2.82$ & $1.60$ & $-$ \\
Differentiable FVN 
& $6.49$ & $.112$ & $3.20$ & $.196$ & $7.85$ & $2.35$ & $-$ \\

\midrule
\multicolumn{8}{r}{\emph{Acoustic matching results}} \\
\midrule

AudioLM-like
& $2.81$ & $.124$ & $3.65$ & $.157$ & $3.47$ & $1.85$ & $-$\\
\quad $-$ Same speaker 
& $2.76$ & $.124$ & $3.60$ & $.157$ & $3.03$ & $1.75$ & $70.9_{\pm 6.0}$ \\
\quad $-$ Different speaker \,\!
& $2.85$ & $.125$ & $3.70$ & $.158$ & $3.74$ & $1.87$ & $66.2_{\pm 6.3}$ \\
%\quad $-$ Blind estimation 
%& $3.65$ & $.209$ & $5.45$ & $.165$ & $2.87$ & $1.56$ \\
RQ-Transformer-like
& $3.05$ & $.120$ & $4.23$ & $.168$ & $2.96$ & $1.79$ & $70.4_{\pm 5.1}$\\
VALL-E-like
& $3.06$ & $.123$ & $4.57$ & $.173$ & $3.39$ & $1.86$ & $66.1_{\pm 6.0}$\\
Differentiable FVN 
& $8.02$ & $.115$ & $3.69$ & $.241$ & $8.16$ & $2.79$ & $57.3_{\pm 5.3}$\\
Non-AR, discrete 
& $3.69$ & $.133$ & $5.08$ & $.202$ & $5.45$ & $4.53$ & $45.3_{\pm 6.6}$\\
Non-AR, continuous 
& $4.24$ & $.132$ & $5.33$ & $.275$ & $10.1$ & $9.73$ & $-$ \\
%%%%%%%%%%%%%%%%%%%%%%%%%%%%%%%%%%%%%%%%%%%%%%%%%%%%%%%%%%%%%%%%%%%
\bottomrule
\end{tabular}
\begin{tabular}{l}
\vspace{-1mm}\\
* LSD, EDR, DRR and $C_{50}$ are mean absolute error (MAE) in \si{dB}. \\
* We report $T_{30}$ error in terms of ratio. \\
* The sujective scores are mean values with $95\%$ confidnece intervals.
\end{tabular}

\vspace{-1.5mm}
\caption{
Evaluation results of the proposed methods and baselines. 
}
\vspace{-9mm}
\label{table:results}
\end{center}

\end{table}

\noindent \textbf{Baselines.} One natural question is whether autoregressive modeling and discretization contributed to the estimation performance. To answer this, we trained a backbone of the RQ-Transformer-like model to predict all indices simultaneously by removing the causal attention mask (Non-AR, discrete). This setup reports clearly degraded performance. We also trained a non-causal TF-Transformer to estimate the continuous latent from the autoencoder (Non-AR, continuous); the results are even worse, especially in terms of DRR and $C_{50}$ error. We also attempted the autoregressive model with continuous features, but the training was unstable. Therefore, discretization and autoregressive modeling were essential for performance under the current framework.
Furthermore, we attached and trained our reference encoder with the differentiable filtered velvet noise \cite{lee2022differentiable, valimaki17fvn}, which models late reverberation with sparse noise segments, each filtered with independent digital filters.
While it showed good MSS loss as the model was optimized with it, it was poor at LSD, which is due to the limited expressive power of the generator. It also showed high DRR and $C_{50}$ errors, similar to the other baselines.
In terms of the SUBJ scores, the differences between the autoregressive models and the non-autoregressive baselines were statistically significant (all $p<1.67\times10^{-3}$ under the pairwise paired t-tests with Bon-Ferroni correction). 
We speculate that the feedforward (all-at-once) estimation and regression objective might have negative impacts on the transient and onset-related performance, but further experiments are required to validate this.

\vspace{-2mm}
\section{Conclusion}
\vspace{-1.5mm}
We explored the two-stage framework for RIR generation and tackled various RIR estimation tasks via conditioned autoregressive decoding. We first obtained the discrete time-frequency token space of RIRs with RQ-VAE, then decoded the token indices with the transformer variants. 
The evaluation results showed that our models estimate perceptually similar RIRs in various tasks, including acoustic matching.  
We also found that both quantization and autoregressive modeling are beneficial for performance; yet, \emph{how much} autoregression yields optimal performance-efficiency tradeoff is still unclear.
Therefore, future works should include improving the tradeoff with lightweight models and optimized autoregressive modeling strategies. Extending the current acoustic matching task to acoustic scene transfer \cite{su20acousticmatch, koo21revconv}, which requires (implicit) signal enhancement, e.g., dereverberation, could be another research direction.

\bibliographystyle{IEEEtran}

\vspace{-1mm}
\newpage
\bibliography{refs23_abbv}
\end{sloppy}
\end{document}